\documentclass[letterpaper]{article} % DO NOT CHANGE THIS
\usepackage{aaai24}  % DO NOT CHANGE THIS
\usepackage{times}  % DO NOT CHANGE THIS
\usepackage{helvet}  % DO NOT CHANGE THIS
\usepackage{courier}  % DO NOT CHANGE THIS
\usepackage[hyphens]{url}  % DO NOT CHANGE THIS
\usepackage{graphicx} % DO NOT CHANGE THIS
\usepackage{caption} % DO NOT CHANGE THIS AND DO NOT ADD ANY OPTIONS TO IT
\usepackage{algorithm}
\usepackage{algorithmic}

\usepackage{newfloat}
\usepackage{listings}
\DeclareCaptionStyle{ruled}{labelfont=normalfont,labelsep=colon,strut=off} % DO NOT CHANGE THIS
\floatstyle{ruled}
\newfloat{listing}{tb}{lst}{}
\floatname{listing}{Listing}
\nocopyright % -- Your paper will not be published if you use this command
\title{Comparison of Encryption Algorithms for Wearable Devices in IoT Systems}
\author{
    Haobo Yang\\
    Mohamed bin Zayed University of Artificial Intelligence\\
    haobo.yang@mbzuai.ac.ae 
}

\usepackage[style=ieee]{biblatex}
\begin{document}

\maketitle
% \tableofcontents

% \begin{abstract}
%     This paper provides an in-depth analysis of encryption algorithms tailored for wearable devices in Internet of Things (IoT) ecosystems. Given the unique challenges associated with wearables, such as limited computational capabilities and the sensitive nature of the data they collect, the paper evaluates a range of symmetric and asymmetric encryption methods in terms of their security, efficiency, and applicability. The paper also explores specialized encryption methods like Format-Preserving Encryption (FPE) and Homomorphic Encryption (HE), assessing their applications in the wearable domain. Through a comparative analysis, the paper aims to guide the selection of encryption methods that meet current security requirements and offer future-proof solutions for the rapidly evolving landscape of wearable IoT devices.
% \end{abstract}

\section{Introduction}

The Internet of Things (IoT) expansion has brought a new era of connected devices, including wearable devices like smartwatches and medical monitors, that are becoming integral parts of our daily lives. These devices not only offer innovative functionalities but also generate and transmit plenty of sensitive data, making their security and privacy the primary concerns. Given the unique challenges posed by wearable devices, such as limited computational resources and the need for real-time data processing, encryption stands as a cornerstone for safeguarding the integrity and confidentiality of the data they handle. Various encryption algorithms, each with its own set of advantages and limitations, are available to meet the diverse security and computational needs of wearable IoT devices. As we move into an age where quantum computing could potentially disrupt traditional encryption methods, choosing a suitable encryption algorithm becomes even more critical. This paper explores and evaluates the suitability of different encryption methods in the context of wearable IoT devices, considering current and future security challenges.

\subsection{IoT and Wearable Devices}

The Internet of Things (IoT) contains a vast network of interconnected devices that serve various purposes, ranging from environmental monitoring \cite{shindeEnvironmentMonitoringSystem2018} to consumer electronics \cite{yunDeviceSoftwarePlatform2015}. Among these devices, wearable technologies, such as smartwatches and fitness trackers, have gained significant popularity. They offer functionalities such as health monitoring, fitness tracking, and convenient access to information, enhancing the quality of life for users \cite{yenSmartWearableDevices2021}.

\subsection{Security Concerns in IoT}

The widespread adoption of IoT devices has raised substantial concerns regarding data security and privacy \cite{weberInternetThingsNew2010}. These devices frequently collect sensitive information, including personal health data and location information \cite{yenSmartWearableDevices2021}. Without robust protection mechanisms, such data can be vulnerable to interception and unauthorized access, posing severe risks to user privacy and security.

\subsection{Role of Encryption}

Encryption is a fundamental security measure in safeguarding the confidentiality and integrity of data within IoT systems. It employs cryptographic algorithms to convert plaintext data into ciphertext data, rendering it unusable to unauthorized parties \cite{mousaviSecurityInternetThings2021}. In this paper, we comprehensively analyse various encryption algorithms suitable for securing data on wearable devices within IoT systems. We compare different encryption methods, both symmetric and asymmetric, to assess their suitability, security, and efficiency in protecting sensitive data on wearable devices.

\section{Encryption Algorithms}

In this section, we provide a general overview of symmetric and asymmetric encryption methods for protecting sensitive data.

\subsection{Symmetric Encryption}
Symmetric encryption, or private-key encryption, uses a single key for encryption and decryption \cite{mousaviSecurityInternetThings2021}. This key is shared between the sender and receiver. Here are some relevant symmetric encryption algorithms:

\begin{itemize}    
    \item \textbf{Triple Data Encryption Standard (3DES):} DES is an outdated symmetric encryption standard with a 56-bit key length, which is no longer considered secure for protecting sensitive data due to technological advances \cite{WhatEncryptionHow}. 3DES is a more secure version of the original DES algorithm, applying DES encryption three times in succession.
    
    \item \textbf{Advanced Encryption Standard (AES):} AES is considered the gold standard for symmetric encryption and is widely used globally \cite{WhatEncryptionHow}. It encrypts 128-bit data blocks at a time and is suitable for various applications, including file and application encryption, WiFi security, VPNs, and SSL/TLS protocols. AES is highly regarded for its security and efficiency \cite{mousaviSecurityInternetThings2021}.
    
    \item \textbf{Twofish:} Twofish is known for its flexibility, speed, and resilience. It operates on 64-bit blocks and is publicly available \cite{SchneierSecurityTwofish}, making it widely accessible. Twofish is suitable for securing e-commerce platforms, password management systems, and email data encryption.
\end{itemize}

\subsection{Asymmetric Encryption}
Asymmetric encryption, also known as public-key encryption, uses a pair of keys: a public key for encryption and a private key for decryption. The public key is freely available, while the private key is kept secret. Below are some well-known asymmetric encryption algorithms:

\begin{itemize}
    \item \textbf{Rivest-Shamir-Adleman (RSA):} RSA is an asymmetric encryption algorithm that leverages the mathematical challenge of factoring large composite numbers to ensure security, where the difficulty in deriving the private key from the public key underpins the robustness of the encryption \cite{mousaviSecurityInternetThings2021}. It is ideal for secure data transmission and identity verification.
    
    \item \textbf{Elliptic Curve Cryptography (ECC):} ECC is considered the next generation of cryptography, using the mathematics behind elliptic curves \cite{mousaviSecurityInternetThings2021}. It offers high security and is difficult to crack due to the mathematical problem it's based on, making it more secure than first-generation systems like RSA.
\end{itemize}

\subsection{Other Encryption Methods}
Apart from the conventional symmetric and asymmetric encryption algorithms, there are specialized encryption methods designed to meet specific requirements. These methods often serve as additional layers of security, built on top of the foundational symmetric or asymmetric algorithms. Some of these are:

\begin{itemize}
    \item \textbf{Format-Preserving Encryption (FPE):} Contrary to the common misconception, FPE is not strictly a symmetric or asymmetric encryption algorithm by itself. Instead, it serves as an additional layer that can be applied on top of either symmetric or asymmetric algorithms . FPE retains the format and length of the data during encryption, making it particularly useful for securing structured data \cite{bellareFormatPreservingEncryption2009}. For example, it can transform a phone number like 012-345-6789 into a ciphertext with a similar format but different, randomized numbers (e.g., 313-429-5072) \cite{DataEncryptionMethods}. FPE is commonly used in cloud management software and tools, making it suitable for cloud data encryption on trusted platforms like Google Cloud and AWS \cite{DataEncryptionMethods}.
    
    \item \textbf{Homomorphic Encryption (HE):} This encryption method allows computations to be performed on encrypted data without requiring decryption first\cite{munjalSystematicReviewHomomorphic2022}. The result, when decrypted, is the same as if the computation had been performed on the plaintext. This is particularly useful for secure data analysis and privacy-preserving computations.
    
    \item \textbf{Quantum Key Distribution (QKD):} As quantum computing poses a threat to classical encryption algorithms, QKD offers a method for secure communication that is theoretically resistant to quantum attacks \cite{sasakiQuantumKeyDistribution2018}. It uses quantum properties to secure a communication channel and is often used in conjunction with other encryption algorithms for added security.
\end{itemize}

\section{General Comparison}
When choosing an encryption method for wearable devices in IoT systems, it is essential to consider several factors that impact the efficiency of the encryption solution. Below, we compare different encryption methods, considering their complexity and security.

\subsection{Complexity}
To compare the complexity of encryption methods, we consider factors such as key generation, encryption, and decryption times. Table~\ref{tab:complexity_big_o} summarizes the time complexity of each method.

\begin{table}[h]
    \centering
    \begin{tabular}{l|c|c|c}
        \hline
        Method & Key Generation & Encryption & Decryption \\
        \hline
        \multicolumn{4}{l}{\textbf{Symmetric}} \\
        \hline
        3DES & $O(1) * 3$  & $O(n) * 3$   & $O(n) * 3$ \\
        AES & $O(1)$ & $O(n)$ & $O(n)$ \\
        Twofish & $O(n)$ & $O(n)$ & $O(n)$ \\
        \hline
        \multicolumn{4}{l}{\textbf{Asymmetric}} \\
        \hline
        RSA & $O(n^4)$ & $O(n^2)$ & $O(n^3)$ \\
        ECC & $O(n^3)$ & $O(n^2)$ & $O(n^2)$ \\
        \hline
        \multicolumn{4}{l}{\textbf{Other}} \\
        \hline
        FPE & \multicolumn{3}{c}{same as underlying algorithm} \\
        HE & N/A & \multicolumn{2}{c}{varies} \\
        QKD & N/A & \multicolumn{2}{c}{varies} \\
        \hline
    \end{tabular}
    \caption{Time Complexity Comparison (Big O Notation), where n is the number of bits of input data \cite{RSALaboratoriesHow, mousaviSecurityInternetThings2021, odeluExpressiveCPABEScheme2017}.} 
    \label{tab:complexity_big_o}
\end{table}

In the realm of symmetric algorithms, DES is viewed as a legacy method for data encryption. The 3DES variant repeats the encryption process three times to bolster its security. While this approach offers improved protection, it comes at the cost of efficiency, making it less optimal than other methods. On the other hand, asymmetric algorithms like RSA generally exhibit higher computational complexity than their symmetric counterparts. Despite providing enhanced security, these algorithms are less efficient regarding computational speed.

Twofish, another symmetric algorithm, has an essential generation process that is slower compared to some other algorithms. This is attributed to its multiple rounds of key mixing and permutation operations, resulting in a linear time complexity with respect to the key size \cite{schneierTwofish128BitBlock}. While this may not be a significant issue for certain applications, it is a factor to consider when evaluating the overall efficiency and suitability of an encryption algorithm for a specific use case. % TODO Twofish: A 128-Bit Block Cipher

\subsection{Security}
Security is of utmost importance when selecting an encryption method. Evaluate the resistance of each method to various types of attacks, including brute force, differential, and known-plaintext attacks.

\begin{itemize}
    \item \textbf{3DES:} DES offers decent security for everyday purposes. However, its primary weakness is its short 56-bit key length, making it vulnerable to brute-force attacks \cite{BruteForce2005}. 3DES, as an improvement over DES, uses a more secure triple-key structure, offering enhanced security compared to DES. However, it is still limited by its 56-bit key length.

    \item \textbf{AES:} AES offers key lengths of 128, 192, or 256 bits, longer than DES methods, providing robust protection against attacks. 

    \item \textbf{Twofish:} Twofish is a flexible and resilient symmetric encryption algorithm suitable for various applications. While it may not be as widely adopted as AES \cite{EverythingYouNeed}, it offers good security.

    \item \textbf{RSA:} RSA is widely supported in various applications. However, its security is contingent on the size of the key used, and smaller key sizes are increasingly considered insecure due to advances in computational power.

    \item \textbf{ECC:} ECC is known for its efficiency and robustness. Additionally, it provides the same level of security as RSA but with much shorter key lengths, making it more efficient for devices with limited computational resources \cite{mousaviSecurityInternetThings2021}.

    \item \textbf{FPE:} The security of FPE also hinges on the strength of the underlying encryption algorithm and the secrecy of the encryption key used \cite{EncryptingDataRestricted}. In other words, it depends on proper implementation.

    \item \textbf{HE:} Homomorphic Encryption allows for computations on encrypted data, making it ideal for privacy-preserving data analysis. However, most HE algorithm is computationally intensive \cite{reisComputinginMemoryPerformanceEnergyEfficient2020} and may not be suitable for all applications.

    \item \textbf{QKD:} Quantum Key Distribution offers a theoretically secure method of communication that is resistant to quantum attacks. It is often used in conjunction with other encryption algorithms for added security \cite{minkQuantumKeyDistribution2009}.
\end{itemize}

In summary, symmetric algorithms like 3DES, AES, and Twofish are faster but require secure key exchange, posing a security risk. Asymmetric algorithms like RSA and ECC eliminate the need for key exchange but are slower and computationally intensive. The choice between the two should be guided by the application's specific security and computational needs.

\section{Discussion}

In this section, we will delve into our topic of how various encryption methods differ from each other in the context of wearable devices.

\subsection{Constraints in Wearable Devices}
When considering the implementation of encryption methods in wearable devices within IoT systems, evaluating the constraints imposed by the device's hardware and software capabilities is crucial. These constraints primarily include key length and memory cost, which directly impact the feasibility and effectiveness of the encryption method.

\subsubsection{Key Length Constraints}
The key length of an encryption method is a vital constraint in the context of wearable IoT devices. Due to the computational limitations of these devices, the key length provides an intuitive insight into the achievable level of security. In Table~\ref{tab:constraints}, we categorize encryption methods based on the feasibility of their key lengths for wearable devices:

\begin{itemize}
    \item \textbf{Short-Key:} Methods like DES and 3DES fall under this category. These methods have shorter key lengths, 56 bits, making them computationally less demanding but less secure.
    
    \item \textbf{Medium-Key:} Encryption methods such as AES and Twofish, which offer key lengths ranging from 128 to 256 bits, are more secure but may require more computational power.
    
    \item \textbf{Long-Key:} Methods like RSA and ECC have long key lengths, providing robust security but may be computationally infeasible for resource-constrained wearable devices.
\end{itemize}

The constraint imposed by key length affects the range of encryption methods that can be feasibly implemented, as longer key lengths may be computationally prohibitive for wearable devices.

\subsubsection{Memory Cost Constraints}
Memory cost is another significant constraint when implementing encryption methods in wearable devices. Given the limited memory resources of these devices, the memory cost of an encryption method can be a deciding factor. In Table~\ref{tab:constraints}, we categorize encryption methods based on their memory cost:

\begin{itemize}
    \item \textbf{Low-Cost:} Methods in this category are optimized for low memory usage, making them suitable for wearable devices with limited resources.
    
    \item \textbf{Moderate-Cost:} These methods require a moderate amount of memory, and their feasibility depends on the specific memory capacity of the wearable device.
    
    \item \textbf{High-Cost:} Methods with high memory demands are generally not suitable for wearable devices, as they could exhaust the device's limited memory resources.
\end{itemize}

The memory cost constraint is critical, as it impacts the device's performance and energy consumption, thereby affecting its operational longevity.

\begin{table}[h]
    \centering
    \begin{tabular}{l|l|l}
        \hline
        \textbf{Method} & \textbf{Key Length (bits)} & \textbf{Memory} \\
        \hline
        \multicolumn{3}{l}{\textbf{Symmetric}} \\
        \hline
        3DES & Short 56 & Low $O(1)$ \\
        AES & Medium 128, 192, 256 & Moderate $O(1)$ \\
        Twofish & Medium 128, 192, 256 & Moderate $O(1)$ \\
        \hline
        \multicolumn{3}{l}{\textbf{Asymmetric}} \\
        \hline
        RSA & Long 1024, 2048, 4096 & High $O(1)$ \\
        ECC & Long 256, 384, 521 & Moderate $O(1)$ \\
        \hline
        \multicolumn{3}{l}{\textbf{Other}} \\
        \hline
        FPE & \multicolumn{2}{l}{\textbf{same as underlying algorithm}} \\ 
        HE & N/A & varies \\
        QKD & N/A & varies \\
        \hline
    \end{tabular}
    \caption{Constraints of Encryption Methods for Wearable Devices in IoT Systems \cite{ mousaviSecurityInternetThings2021, DataEncryptionMethods}}
    \label{tab:constraints}
\end{table}

\subsection{FPE for IoT Application}
Format-Preserving Encryption (FPE) is an additional layer on top of other encryption methods, for example, ECC. Its unique capabilities make it a compelling choice for wearable IoT devices for several reasons:

In the evolving landscape of IoT security, wearable devices present unique challenges due to their limited computational capabilities and the sensitive nature of the data they collect. Traditional encryption methods often struggle to balance the need for strong security with the constraints of real-time data processing and limited resources. This is where FPE comes into play, offering a tailored solution that addresses these specific challenges.

\begin{itemize}
    \item \textbf{Preservation of Data Format:} FPE maintains the original data format, which is particularly useful for wearable devices that require structured data storage or transmission. This feature simplifies data management and ensures compatibility with existing systems that expect data in a specific format.
    
    \item \textbf{Modular Encryption:} FPE divides the encryption tasks into separate parts, allowing for more flexible and efficient encryption processes. This is beneficial for wearable devices with limited computational resources.
    
    \item \textbf{Enhanced Security:} By serving as an additional layer above other encryption methods, FPE can increase overall data security. This is crucial for wearable devices that collect sensitive information, such as health metrics.
\end{itemize}

The advantages of FPE are not just theoretical but have practical implications for the design and operation of wearable IoT devices. By adopting FPE, manufacturers can achieve a higher level of security without compromising on performance or compatibility. This makes FPE a future-proof choice, especially considering the looming threats posed by advancements in quantum computing.

\begin{itemize}
    \item \textbf{Algorithm Agnosticism:} FPE's security is dependent on the underlying encryption algorithm. This means that as long as the underlying algorithm (e.g., ECC) is secure and possibly quantum-resistant, FPE will inherit these security properties.
    
    \item \textbf{Real-time Performance:} Due to its modular nature and the ability to work with efficient underlying algorithms, FPE can meet the real-time data processing requirements commonly found in wearable IoT devices.
    
    \item \textbf{Regulatory Compliance:} FPE can be configured to meet various regulatory requirements for data protection, such as HIPAA for medical data, making it a versatile choice for different types of wearable devices.
\end{itemize}

In summary, FPE's unique features, such as format preservation, modular encryption, and the ability to layer on top of other secure algorithms, make it a suitable encryption method for wearable IoT devices. Its flexibility and enhanced security measures can help these devices meet both functional and security requirements, making it a robust choice for current and future applications.

\subsection{Quantum Attack Challenge}
With the advent of quantum computing, traditional encryption methods, including those used in wearable devices, face new challenges. Quantum computers have the potential to perform certain types of calculations much faster than classical computers \cite{laddQuantumComputers2010}, posing a threat to the security of existing encryption schemes. This subsection explores the quantum attack challenge and assesses the vulnerability of various encryption methods commonly used in wearable devices.

\subsubsection{Vulnerable to Quantum Attacks:}
Here, we highlight encryption methods that are vulnerable to quantum attacks:

\begin{itemize}
    \item \textbf{DES:} DES is vulnerable to quantum attacks due to its short 56-bit key length, making it susceptible to brute-force attacks when quantum computers become powerful enough.
    
    \item \textbf{3DES:} Similar to DES, 3DES is also vulnerable to quantum attacks, as it retains a limited key length of 56 bits.
    
    \item \textbf{AES:} AES can be vulnerable to quantum attacks when used with key lengths less than 256 bits.
    
    \item \textbf{Twofish:} Twofish, like AES, is vulnerable to quantum computing vulnerabilities.
    
    \item \textbf{RSA:} RSA is vulnerable to quantum attacks when quantum computers reach sufficient computational power.
    
    \item \textbf{ECC:} ECC is also vulnerable to quantum computing vulnerabilities.
\end{itemize}

\subsubsection{Not Vulnerable to Quantum Attacks:}
On the other hand, some encryption methods are not inherently vulnerable to quantum attacks:

\begin{itemize}
    \item \textbf{FPE:} FPE is not inherently vulnerable to quantum attacks; its security depends on proper implementation.
    
    \item \textbf{HE:} Homomorphic Encryption is not inherently vulnerable to quantum attacks, but its computational intensity may limit its applicability in wearable devices.
    
    \item \textbf{QKD:} Quantum Key Distribution is designed to be quantum-resistant and offers a secure method of communication that is theoretically impervious to quantum attacks.
\end{itemize}

In summary, the advent of quantum computing poses a significant challenge to the encryption methods commonly used in wearable devices. As quantum technology advances, it becomes increasingly important to consider quantum resistance when selecting encryption methods for wearable devices.

\subsection{Suitability of Encryption Algorithm}

Wearable devices often have limited computational resources and battery life, making the choice of encryption algorithm crucial for both security and efficiency. Below is an evaluation of the suitability of various encryption algorithms for wearable devices, along with example applications:

\begin{itemize}
    \item \textbf{3DES:} While 3DES is compatible with legacy systems and relatively simple to implement, it is computationally more intensive than modern algorithms. Its 56-bit key length also poses a security risk. It may be suitable for basic fitness trackers that require compatibility with older systems.
    
    \item \textbf{AES:} AES is efficient and offers robust security, making it a preferred choice for advanced smartwatches like the Apple Watch or Samsung Galaxy Watch, which require both efficiency and high security.
    
    \item \textbf{Twofish:} Twofish is similar to AES in terms of security but is publicly available, making it a viable option for open-source wearable projects like open-source smart glasses or DIY fitness trackers.
    
    \item \textbf{RSA:} RSA is computationally intensive and tends to be slower than symmetric algorithms. It could be used in high-security applications like digital identity cards or secure key fobs where computational time is not a critical requirement.
    
    \item \textbf{ECC:} ECC offers high security with shorter key lengths compared to RSA. This makes it more efficient and thus suitable for resource-constrained devices.

    \item \textbf{FPE:} FPE is specialized and may be suitable for medical wearables that require structured data encryption, such as glucose monitors or ECG monitors.
    
    \item \textbf{HE:} Homomorphic Encryption is computationally intensive and may not be suitable for real-time applications in wearable devices. However, it could be used in specialized medical research wearables where data privacy is of utmost importance.
    
    \item \textbf{QKD:} Quantum Key Distribution is theoretically secure against quantum attacks but is currently impractical for wearable devices due to its high computational and hardware requirements.
\end{itemize}

In summary, the choice of an encryption algorithm for wearable devices should balance security needs and computational limits. ECC and AES are generally recommended due to their optimal blend of security and efficiency for most wearable applications.

\section{Conclusion}
In conclusion, selecting an appropriate encryption method for wearable devices in IoT systems is a complex yet crucial task that significantly influences data security and operational efficiency. Given the limited computational resources and battery life commonly associated with wearable devices, the choice of encryption algorithm must be carefully tailored to meet these constraints without compromising security. Algorithms like AES and ECC generally offer a balanced solution, providing robust security while being computationally efficient. Specialized methods like FPE and HE have niche applications, particularly in medical wearables and privacy-sensitive tasks. As quantum computing advances, considerations for quantum-resistant algorithms are increasingly becoming a focal point in decision-making. Therefore, the choice of encryption method should be a well-informed decision, considering the specific needs, capabilities, and future-proofing requirements of the wearable IoT application.

\printbibliography

\end{document}